# Gain stabilization in Micro Pattern Gaseous Detectors: methodology and results

**Dan Shaked Renous[1], Arindam Roy, Amos Breskin and Shikma Bressler**

*Dept. of Astrophysics and Particle Physics, Weizmann Institute of Science,*
*Rehovot, Israel.*
*E-mail:* `dan.shakedrenous@weizmann.ac.il`

ABSTRACT: The phenomenon of avalanche-gain variations over time, particularly in Micro Pattern Gaseous Detectors (MPGD) incorporating insulator materials, have been generally attributed to electric-field modifications resulting from "charging-up" effects of the insulator. A robust methodology for characterization of gain-transients in such detectors is presented. It comprises three guidelines: detector initialization, long-gain stabilization monitoring and imposing transients by applying abrupt changes in operation conditions. Using THWELL and RPWELL detectors, we validated the proposed methodology by assessing a charging-up/charging-down model describing the governing processes of gain stabilization. The results provide a deeper insight into these processes, reflected by different transients upon abrupt variations of detector gain or the irradiation rate. This methodology provides a handle for future investigations of the involved physics phenomena in MPGD detectors comprising insulating components.

---

[1] Corresponding author.

**Contents**

# 1  Introduction

The article proposes a method for investigating in a systematic way charge-multiplication properties in gas-avalanche detectors. The phenomenon of avalanche-gain variations over time is well-known, particularly in Micro Pattern Gaseous Detectors (MPGD) incorporating insulator materials. It has been observed, for example, with Gas Electron Multipliers (GEM) [1-3], THick Gaseous Electron Multipliers (THGEM) [4, 5] and other gas-avalanche amplifying structures [6-8]. These variations have been generally attributed to electric-field modifications resulting from "charging-up" effects of the insulator. In GEM or THGEM, for example, the electrode's insulating substrate (e.g. Polyimide and FR4, respectively) might undergo two competing processes – charge accumulation (*charging up*) and charge evacuation (*charging down*). Regardless of the underlying physics processes contributing to these phenomena, gain stabilization is expected once they reach equilibrium. One process contributing to the charging up is the lateral diffusion of the avalanche charge. It has been suggested that in THGEM electrodes with cylindrical holes, this process results in the avalanche electrons occupying the lower part of the hole walls and the ions occupying its upper one [9]. This charge distribution creates an electric field opposing the original one, thus reducing the gain. In GEM electrodes with double-conical holes, avalanche ions accumulate mostly at the bottom of the conical surface while the electrons accumulate at the top one [3]. Indeed, a rise in gain was witnessed [1], in agreement with the model of charge accumulation. While charging-down processes were neglected due to long typical time scales involved, the condition for gain stabilization in [3] was that after some time, there is a similar probability for additional electron accumulation as additional ion accumulation. Thus, the field created by the additional electrons distribution cancels out the effect of the additional ions distribution.

In the context of MPGDs, fewer studies assessed the question of charge evacuation. Possible mechanisms involved in this process could be recombination of electrons and ions with gas molecules, surface and bulk conductance by humidity and by the electric field applied between the electrodes. A recent charge-clearance model for THGEM-like detectors [10] argues that the charges move within the printed-circuit board (PCB) fiberglass along the electric field lines. This effect, not to be confused with dielectric polarization, is related to the mobility of charges in the bulk. Thus, it is a slow process and an asymptotic gain-stabilization behavior is reached within several hours or days. In detectors with resistive materials, such as the Resistive-Plate WELL (RPWELL [11-14]), gain variations could additionally arise from the presence and properties of the Resistive Plate (RP).

In early studies with THGEM detectors [4, 15], it was shown that gain instabilities with time varied according to the detector's "history" and total avalanche charge. This observation has been systematically revisited in this work, showing that stable gain could be reached in all cases, regardless of the detector's history. However, the time-scale needed for reaching a steady-state, namely equilibrium between the charge *accumulation* and *evacuation,* can differ among the configurations investigated - being dependent on many parameters. For example, the gain measured with an electrode exposed to ambient air (e.g. to humidity) prior to the measurements, stabilizes after a longer time compared to the gain measured with an electrode irradiated after long gas circulation; transient gain-stabilization effects have been observed following abrupt irradiation-flux variations or that of the operation voltage. Therefore, a procedure had to be developed for "initializing" the detector's conditions prior to its investigations, as well as a methodology for measuring the gain stabilization in a consistent way. This permitted reaching reproducible results and defining, in a consistent way, the meanings of the quoted values of gain.

The hypothesis of our study is that the field variations are dominated by the changes in charge distribution on the insulating and resistive materials of the detector. It considers the charging-up and charging-down processes which could originate from different mechanisms. Demonstrated with a Thick WELL (THWELL) [16] and RPWELL [11] configurations, an attempt was made to isolate the two competing processes and to quantify their dominating time scales. The methodology developed here provides a handle for future investigations of the involved physics phenomena in MPGD detectors comprising insulating components.

This paper is organized as follows: our underlying model is presented in section 2; in section 3 we describe the experimental setup; the methodology for gain stabilization studies is discussed in section 3; the results of a comparative study of various detector configurations are presented in section 4, followed by a discussion in section 5 and a summary in section 6.

## 2  The "charging-up/charging-down" model

For simplicity, we start by considering a THWELL configuration, i.e. a single-sided THGEM electrode coupled directly to a readout anode without additional resistive components. The gain ($G$) depends on the first Townsend coefficient ($\alpha$) and the effective electric field along the THGEM hole axis ($E$) in the multiplication region by the following relation:

$$G \sim e^{-\alpha E} \tag{1}$$

We approximate $E$ as follows

$$E = E_0 - E_Q \tag{2}$$

Here, $E_0 = E_0(\Delta V, d^{-1})$ is the field resulting from the detector geometry (in THWELL, $d$ is the electrode thickness) and $\Delta V$ - the difference between the voltage supplied to the different electrodes. In this definition $E_0$ is constant. $E_Q$ is the electric field resulting from the distribution of the charges accumulated on the insulating walls ($Q$). $E_Q$ is subtracted from $E_0$ as a convention since, as discussed earlier, in THGEM detectors this component reduces the total field. Denoting the rates of the charging up/down processes by $\Gamma_{up/down}$ we get:

$$Q(t) \propto \int_0^t [\Gamma_{up}(\tau) - \Gamma_{down}(\tau)]d\tau \qquad (3)$$

According to our hypothesis, gain stabilization occurs at time $t_0$ when for each $t > t_0$, $\Gamma_{up}(t) = \Gamma_{down}(t)$. $\Gamma_{up/down}$ could have contributions from different mechanisms and they have, in general, a complicated structure. For example, the charge up process could be dominated by the lateral diffusion of the avalanche charges. This diffusion is more dominant when the field in the holes, $E$, is smaller [17] or when the charge density in the avalanche – determined by the gain – is large. The field resulting from the charge accumulating on the holes, $E_Q$, also affects the diffusion. Mechanisms that could be involved in the charge down process are recombination of electrons and ions with the gas molecules, surface and bulk conductance, humidity and the electric field applied between the electrodes and more. Both $\Gamma_{up}$ and $\Gamma_{down}$ depends on the gas and the insulating material properties

Under this simplified model, two typical gain stabilization profiles (transients) are identified: charging-up dominated, $\Gamma_{up}(t) \geq \Gamma_{down}(t)$, and charging-down dominated $\Gamma_{up}(t) \leq \Gamma_{down}(t)$. The condition for steady-state ($\forall\ t > t_0$: $\Gamma_{up}(t) = \Gamma_{down}(t)$) means that for every charge that makes it to the walls, there is one that leaves.

The extreme condition, where $\Gamma_{down}(t) = 0$, implies that at steady-state no charge can be accumulated on the walls; in agreement with the model discussed above. In such case, a change in the irradiation rate would not affect the gain at equilibrium. If we consider a constant $\Gamma_{down}$ above zero, $\Gamma_{up}$ at steady-state would be non-zero inferring a finite probability for charge accumulation. Hence, by increasing $\Gamma_{event}$ or the avalanche size, $\Gamma_{up}$ would also increase, breaking the equilibrium. Thus, a larger $Q$ will be needed to reduce $\Gamma_{up}$ once more to reach equilibrium.

In an RPWELL configuration, where the THGEM is coupled to the readout anode via a plate of high resistivity material (RP), $E_0$ is no longer constant. It is affected by possible voltage drops on the RP due to the current (I) flowing through it, $\Delta V \to \Delta V - IR$; where $I$ and $R$ can be approximated by the expressions:

$$I \sim N_{PE} q_e G \Gamma_{event} \qquad (8)$$

$$R = \rho \cdot \frac{d_{RP}}{A_{RP}} \qquad (9)$$

Here $N_{PE}$ is the number of primary electrons, $q_e$ is the charge of the electron, $\rho, d_{RP}$ and $A_{RP}$ are the bulk resistivity, the thickness and the effective area of the RP, respectively. Thus, the gain is affected by the presence of the RP [18]. An increase of $\Gamma_{event}$ or avalanche size will be translated into a decrease of $E_0$.

The main objective of the gain-stabilization study presented in this work has been to decouple, as much as possible, charging-up dominated regions from ones dominated by charging down. This was done by exposing the detector to rapid transitions in the operation conditions – operation voltage and irradiation rate.

## 3    Experimental setup

The experimental setup is shown in figure 1. The investigated detectors were placed in an aluminum vessel flushed with Ne/5%CH$_4$ at a flow of 20 sccm; the vessel could be pumped down to high vacuum. The detectors were irradiated with 8 keV photons from an x-ray tube (Cu-target)[2], illuminating the detector through a 50 mm diameter thin Kapton window; the x-ray flux was controlled by varying the tube's current and by a stack of thin Cu filters placed before the Kapton window. Power to the drift electrode and the multiplier's top electrode was supplied through a low-pass filter using CAEN HV A1651H module and SY5527 mainframe; the detector's current and voltage monitoring was done using CAEN GECO2020 software. The signals were recorded from the anode through an Ortec 125 charge-sensitive preamplifier; they were processed either by an Agilent Technologies InfiniiVision DSO-X 3034A oscilloscope - for signal-shape acquisition, or by an Amptek[3] 8000a pocket Multi-Channel Analyzer (MCA), through an Ortec 572 linear amplifier - for pulse-height spectra acquisition.

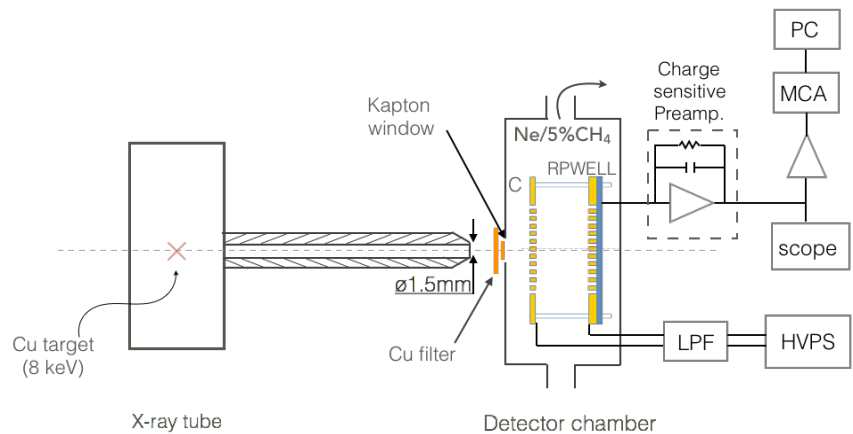

**Figure 1.** Experimental setup: The detectors (shown is a RPWELL) placed in a vessel, under gas flow, are irradiated by a collimated 8 keV x-ray beam. Signals from the detector's anode are –recorded with a charge-sensitive preamplifier, shaped, and processed by a digital oscilloscope or a multi-channel analyzer (MCA). The detector electrodes voltage is supplied by the high-voltage power supply (HVPS) through a low-pass filter (LPF).

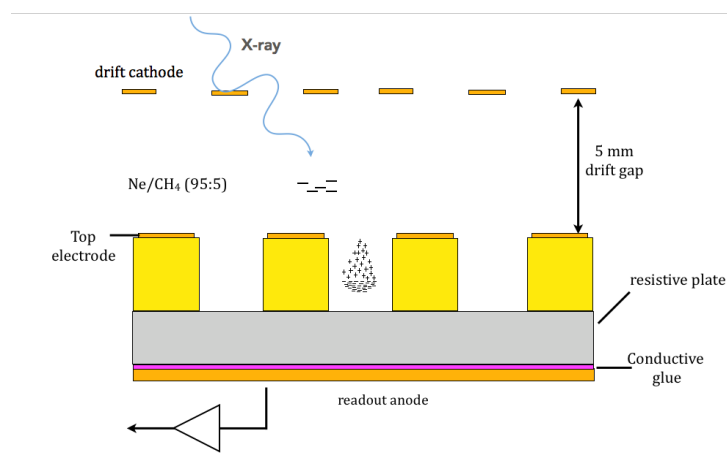

---

[2] Oxford Instruments Apogee 93500 radiation shielded tube
[3] http://amptek.com

**Figure 2.** The RPWELL detector configuration. X-ray induced electrons are collected into the holes, undergoing avalanche multiplication; avalanche charges induce signals on the Cu anode located beneath a resistive plate. The THWELL detector has a similar configuration, but with the multiplier directly coupled to the Cu anode.

Two types of MPGD detectors were investigated for validating the proposed methodology: 1) a 0.4 mm thick, bare THWELL (denoted **0.4-THWELL**) detector [19] – a single-sided THGEM electrode, copper clad on one side, coupled directly to a copper anode; 2) 0.4 and 0.8 mm thick RPWELL (denoted **0.4-RPWELL** and **0.8-RPWELL**) detectors – single-sided THGEM electrode, copper clad on one side, coupled to the anode through a 0.4 mm thick resistive plate (RP) of high bulk resistivity (here, Semitron ESD125; $\sim 10^9$ $\Omega$cm bulk resistivity). Figure 2 depicts a typical RPWELL detector configuration; note that some preliminary detector investigations (figure 3) were done with a "glass-RPWELL", using a LRS-glass RP of $10^{10}$ $\Omega$cm bulk resistivity [20].

## 4 Methodology

The methodology developed within this work targets at assessing the question of gain stabilization through its response to abrupt transition in the operation conditions. It involves three main ingredients: 1) Initialization of the detector history 2) Reaching gain stabilization and 3) Imposing transients. The methods described below resulted in a consistent and a reproducible set of results.

### 4.1 Detector initialization

Each pre-used electrode, that has previously been exposed to radiation-induced avalanche multiplication, was found to perform differently - thus required "initialization" prior to the detector assembly. The method we found both adequate and fast involved *neutralizing* the pre-accumulated charges on the electrode by its exposure to ambient air and its flushing for ~1 minute with a jet of ionized nitrogen; this was performed with an anti-static gun (Electrostatics Inc. model190M). Once "electrically clean", the detector was assembled inside the test chamber and flushed with Ne/5%CH$_4$ at a flow-rate of 20 sccm for four days to remove humidity. An equal result has been reached by pumping the detector for 24 hours, to high vacuum.

### 4.2 Gain stabilization: long-term measurements

The key-principle in detector-performance evaluation is waiting for gain stabilization after each modification of the operation conditions – followed by a few-hours waiting-period to assess stability. Careful systematic measurements have indicated that a few effects can be involved in the gain-stabilization process; each could be governed by a different time scale. An example is depicted in figure 3. Two hours after setting a potential of 700 V across a 0.8 mm thick glass-RPWELL, the gain seemed having reached stability; in fact, it only reached a minimum value (as explained below in section 5).

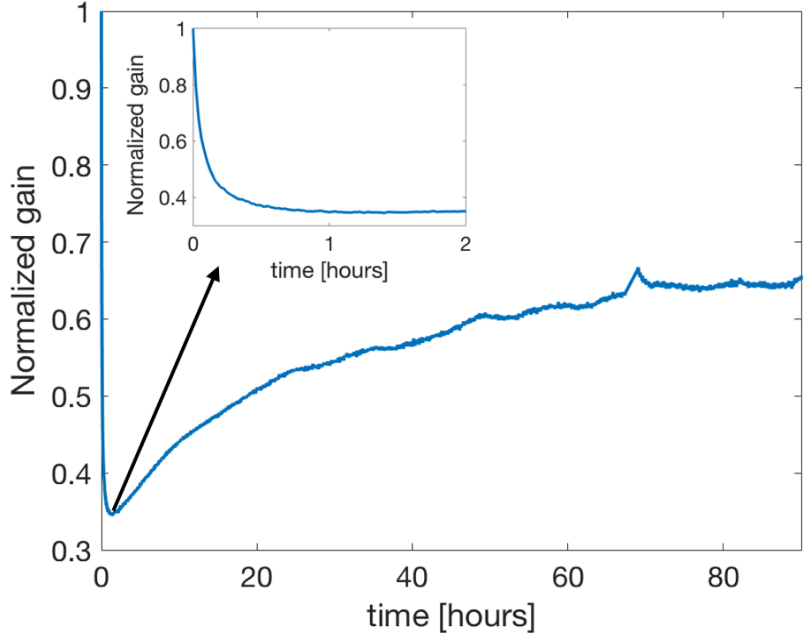

**Figure 3.** Gain as a function of time of a RPWELL detector (of figure 2) having a 0.8 mm thick FR4-THGEM electrode coupled to a 0.7 mm thick LRS glass RP. Ne/5%CH$_4$; 8 keV photons at 30 Hz/mm$^2$. (a) gain evolution for the first 2 hours and (b) gain evolution over 80 hours.

### 4.3 Imposing transients

The detector response to transitions in its operation conditions was studied under abrupt changes of different parameters (so-called stress-tests). Two kinds of transitions were investigated: varying either the operation voltage (gain) or the irradiation-rate. Detectors were first investigated under an order-of-magnitude gain variation (from $10^3$ to $10^4$ and back to a gain of $10^3$). Then, they were exposed to several cycles of close to two orders-of-magnitude changes in the irradiation rate (from 30 Hz/mm$^2$ to 1.3 kHz/mm$^2$ and vice-versa). As discussed above, for each introduced transition, the gain was monitored – till its stabilization.

In view of the charging up/down model discussed in section 2 the following gain variation profiles are foreseen: Once the voltage is increased, $E_0$ (and hence the gain) increased promptly. We expect $\Gamma_{up} > \Gamma_{down}$, i.e. a charging-up dominated transient. In THWELL, comprising cylindrical holes, this would result in a steady-state at a lower effective gain than the one set by the operation voltage. Once the voltage is decreased, $E_0$ and hence the gain decreased promptly. At this point, a considerable amount of charge ($Q$) is already accumulated on the insulators, more than the one expected at the steady-state at that voltage. In THWELL, this would result in field and gain which are lower than those expected at steady-state. Hence, the fast gain drop will be followed by an increase of the gain ($\Gamma_{down} > \Gamma_{up}$).

An abrupt increase of the irradiation rate is expected to result in more charge accumulating in the holes and hence in charging-up dominated profile. A sharp decrease of the irradiation rate will reduce the amount of incoming charge and a charging-down process will dominate. In addition to the charging up/down effects, the change of the irradiation rate modifies $E_0$; at high rates, we anticipate voltage drops, due to the currents flowing through the resistive plate. Such rate-dependence of the gain is supported by the results of past studies [11, 16].

Results below are presented in terms of the average gain, defined as the mean value of a Gaussian fit to pulse-height spectra measured over 1 minute; this allowed monitoring gain variations over short time periods.

## 5   Results

Qualitative results with the 0.4-RPWELL figure 4. shows a complete sequence of transient measurements. In this example, the results are shown for a 0.4-RPWELL detector (of figure 2). Note that, qualitatively, the observed properties were common to all the configurations investigated. The measurement demonstrates the transients in gain, associated with the abrupt changes made in the multiplier's voltage and in the irradiation rate. Time-zero indicates the moment of activating the x-ray tube. The measurements are divided into six main time intervals: (a) initial gain stabilization (to a steady-state) at an operation voltage equivalent to a stable low-gain value of ~1000; it is followed by gain stabilization after: (b) transition from the low-to-tenfold-higher gain; (c) returning to the initial voltage value; (d) transition from irradiation rate of 30 Hz/mm$^2$ to that of 1.3 kHz/mm$^2$; (e) transition back to 30 Hz/mm$^2$; and (f) a second cycle of low-high-low irradiation rates.

Several observations are derived from figure 4:

*Gain value at steady-state:* in the transition from low (region a) to high voltage (region b) and then back to the original low voltage (region c), the gain value at steady-state remains unchanged. On the other hand, in the transition from low rate (region c) to high rate (region d) and then back to low rate (region e) the gain didn't return to its original value and a new steady-state, with a gain higher than the original one, was measured. A second cycle of rate increase and decrease (region f) didn't result in a change of the steady-state.

*Transition from low-to-high voltage/gain (region b):* in response to an increase in operation voltage, all the investigated configurations showed fast increase in gain, followed by a drop. This response is in agreement with the described model, specifically the charging up dominance transient.

*Transition from high-to-low voltage/gain (region c):* both RPWELL configurations responded with a fast drop below the nominal stable gain value, followed by a slow increase of the gain up to its full recovery. This result is in agreement with the charging up/down model, where the charge down process is dominating.

*Transition from low-to-high irradiation rate (region d):* all detector configurations responded to the first increase in irradiation rate with a fast drop in gain followed by a slow increase to a steady-state gain value higher than the one measured at low irradiation rate. This result, which was well pronounced also in the THWELL configuration, does not agree with previous observations [11, 16]. In the context of the model discussed in section 2, the indicated gain-drops upon increase of the irradiation rate were attributed to the distribution of the accumulated charge on the cylindrical holes of the THWELL.

*Transition from high-to-low irradiation rate (region e):* resuming the low irradiation rate conditions didn't result with the same gain value obtained prior to irradiation-rate increase. This counter intuitive result indicates a change in the steady-state condition. In light of the model discussed in section 2, such a change could be explained by the modification of the field configuration due to the massive accumulation of charge on the holes ($Q$).

*A second cycle of low-high-low irradiation-rate transitions (region f):* the detectors' response to a second cycle of irradiation-rate transitions yielded the same gain values as in the first cycle.

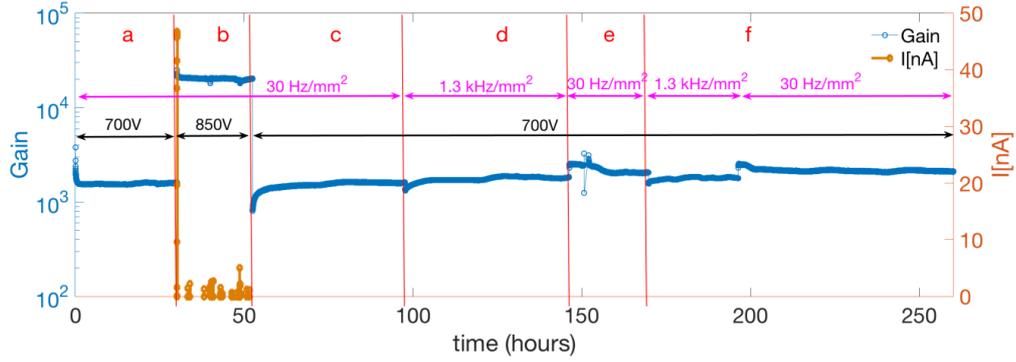

**Figure 4.** Gain vs. time of a 0.4-RPWELL (figure 2); measurements in Ne/CH$_4$ (95:5) with 8 keV photons, started after 4 days of gas flushing. (a) irradiation-rate of 30 Hz/mm$^2$ under V$_{RPWELL}$=700 V; (b) raising V$_{RPWELL}$ to 850 V; (c) reducing V$_{RPWELL}$ to 700 V. (d) at V$_{RPWELL}$ = 700 V, raising the rate to 1.3 kHz/mm$^2$; (e) rate restored to 30Hz/mm$^2$; (f) a second cycle of low-high-low rate transitions. Qualitatively, a similar trend was observed with all the investigated detector configurations.

## 5.1 Comparative results

In the following, we compare the gain-stabilization results of the various detector configurations: 0.4-THWELL, 0.4-RPWELL and 0.8-RPWELL. Three main stages are compared: the initial stabilization stage (figure 5), low-high-low gain transitions (figure 6) and low-high-low rate transitions (figure 7).

Once the source is switched on for the first time, the gain value changes fast. Hence, the measured gain value instantly after switching on the source is subject to large uncertainty. In order to avoid biasing our conclusions, the results presented below are normalized to the gain value measured once the initial steady state was reached (region "a" in figure 4).

### 5.1.1 Initial stabilization (region a in figure 4)

As discussed earlier, all configurations showed initialization profile that matches a charging-up dominated behavior.

Figure 5 compares the effects of two initialization methods on the gain stabilization: an initial 4-days gas flushing to that of evacuating the detector chamber down to 1.5×10$^{-4}$ Torr for 24 hours (with a dry turbo-molecular pump) followed by 24 hours gas flushing. These methods are compared also to a detector which was flushed only for 5 hours – the minimal time required to fill up the detector vessel. While the pumping followed by gas circulation resulted in gain stabilization after ~2h (to about 37% of the initial value), the 4-days gas flushing method resulted in a similar gain stabilization process as the minimally-flushed detector, where a long-term increase of the gain was shown.

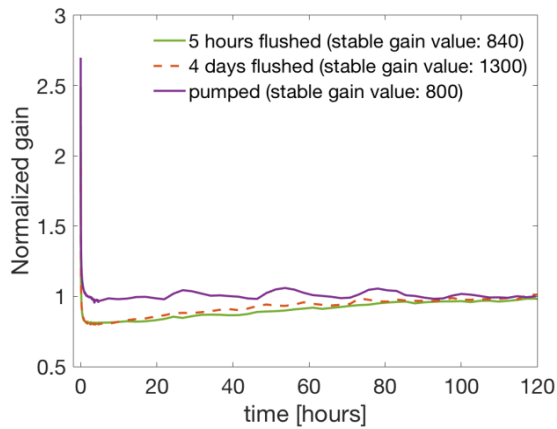

**Figure 5.** Normalized gain stabilization of a 0.8-RPWELL detector, operated at 800 V under a 30Hz/mm² 8 keV x-ray flux. Operation after three initialization methods: (1) control - flushing Ne/CH$_4$ (95:5) for 5 hours; (2) flushing Ne/CH$_4$ (95:5) for 4 days; (3) vacuum of $1.5 \times 10^{-4}$ Torr for 24 hours followed by Ne/CH$_4$ (95:5) flushing for 24 hours prior to measurements.

### 5.1.2 Low-high-low voltage/gain transitions (regions b-c):

Figure 6a depicts the response of the 0.4-RPWELL and 0.8-RPWELL detectors to a tenfold increase in gain (increasing the operation voltage). Both detectors reached gain stabilization within ~30-60 minutes with somewhat shorter stabilization time measured with the 0.4-RPWELL. This could be attributed to the difference in electrode thickness, the slightly different initial gain or to the different electric field, indicating the difficulty to decouple the many parameters governing the charging-up process.

Restoring the voltage to the initial lower value, (figure 6b), both detectors responded with significant initial gain drop (the actual value is attributed to large uncertainties); the gain of the 0.4-RPWELL recovered within ~5h to its initial value; that of the 0.8-RPWELL, within ~15h. Note that the fast gain increase in the 0.4-RPWELL, observed after the 20$^{th}$ hour, is correlated with temperature fluctuations (not corrected for).

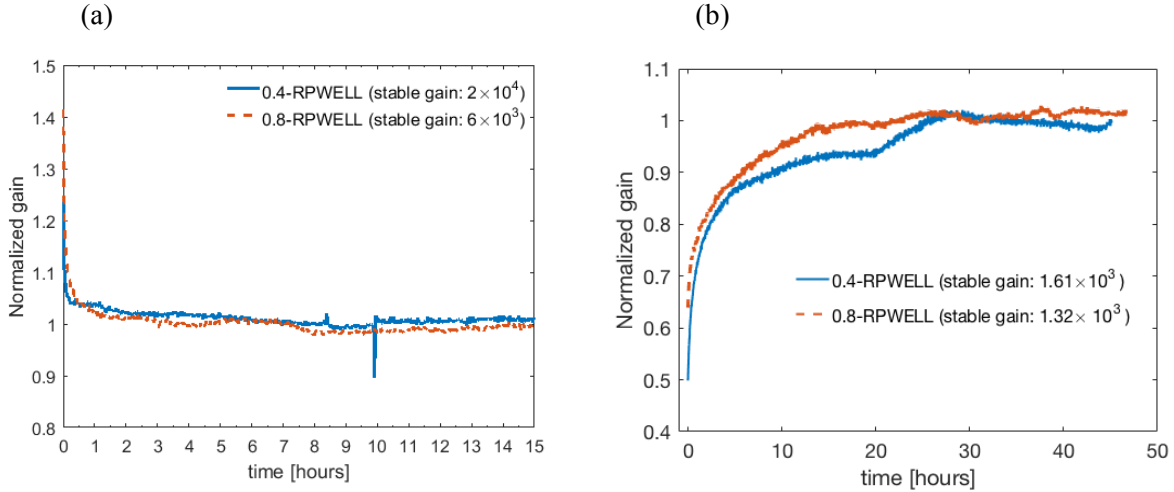

**Figure 6.** Normalized gain variations vs. time for 0.4- and 0.8-RPWELL detectors (of figure 2). (a) voltage increased to yield a tenfold higher gain (values shown in the figure); (b) voltage decreased to its initial low-gain settings. The initial (*the higher*) voltage settings were 700 V (*850V*) and 800 V (*900V*) for the 0.4- and 0.8-RPWELL detectors, respectively. Note that the gain in (a) was normalized to the gain value after 15 hours and in (b) to the stable gain value in region "a" in figure 5 (see text).

### 5.1.3 Rate transitions (regions d-f)

When assessing the response to each rate transition we considered two parameters: the gain 1 minute after rate variation and the stable gain (value after stabilization). Figure 7 depicts the stable-gain value following each rate transition, from 30 Hz/mm² to 1.3 kHz/mm² and vice-versa normalized to the value measured before a given transition. Two significant results are observed:

1) Both RPWELL and THWELL configurations responded to the first transition to a higher rate with a higher stable-gain value compared to that of the lower rate.

2) In the following rate transitions, the RPWELL structures responded to low rate with further gain increase, while the gain of the THWELL detector, stabilized at a 40% higher value (after the initial rate increase), remained unchanged regardless of further successive rate variations.

The second observation agrees with the rate dependency expected by the premise of our study, which expects a voltage rise (*drop*) due to the decrease (*increase*) of current through the RP. Hence the THWELL response, lacking a RP layer, should show less pronounced rate dependency and be dominated by the charging up/down equilibrium of the walls. In the context of the model discussed in section 2, the response of the THWELL indicates that the charge down rate, $\Gamma_{down}$, is small and does not play a significant role in the range of irradiation rates tested in this study. Further discussion will be presented in the next section.

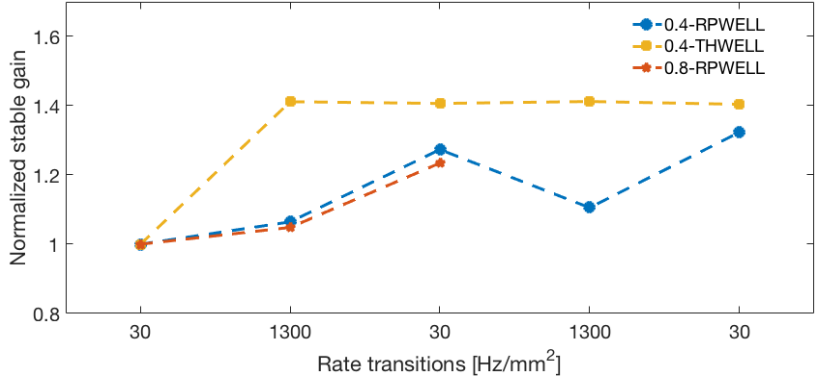

**Figure 7.** The normalized stable-gain values, in Ne/ %5 CH$_4$, of the two RPWELL configurations and the THWELL, following irradiation-rate transitions.

## 6   Summary and discussion

A robust methodology for the characterization of transients in MPGD was established. It is based on a well-defined initialization process and a long-term stabilization period. The method was validated on the Resistive-Plate WELL (RPWELL) and the Thick-WELL (THWELL) detectors, yielding a set of reproducible results that provided a good insight into their response. It can pave the way towards a systematic study of the physics processes governing the response of gas-avalanche detectors, in particular those incorporating insulators and resistive materials.

Using this methodology, the response of the THWELL and RPWELL detectors was studied in view of a suggested charging-up/down model. This model is a preliminary attempt to understand trends in the electric field and gain ($E$ and $G$ respectively) stabilization of these detectors. The model relies on two underline assumptions:

1. The total field is a superposition of two components: the first is $E_0$– the field resulting from the detector geometry and the applied voltage on the electrode. The second is $E_Q$ – the field resulting from the distribution of the charges accumulated on the insulating walls. Field modifications due to the avalanche charges are neglected.
2. The charging up rate, $\Gamma_{up}$, and the charging down rate, $\Gamma_{down}$, are governed by different physical processes and could have different characteristic times.

Abrupt variations of the operation voltage (and gain) and of the irradiation rates were introduced; they affected $E_0$ and $E_Q$ differently, giving rise to different gain stabilization trends. Moreover, the increasing and decreasing of the operation voltage or irradiation rate imposes charge up and down dominated responses respectively. These procedures decouple, as much as possible, the processes governing the gain stabilization of the detector. The main findings are:

*Initial gain stabilization:* In past work with THGEM electrodes [10], the observed phenomenon of gain increase during the initial stabilization process was attributed to the charge movement within the 0.4 mm thick (FR4) THGEM-electrode substrate, under high electric fields. Since the charge displacement within the insulator bulk is claimed to be slow, the stabilization time-scale can be in the order of hours or days. Such phenomenon was not observed with the 0.4-RPWELL investigated in this work.

However, a long-term gain increase was measured with the 0.8-RPWELL (Figure 5) when it was not pumped prior to its operation. Therefore, it can be associated with outgassing of the electrode substrate, that could be affecting the gas purity or the insulator conductivity.

*Stable-gain values:* The same stable-gain value was recorded before and after the low-high-low operation voltage cycle. Different stable-gain values were recorded before and after the first low-high-low irradiation rate cycle. This indicates that the two transitions affect the detector in a different way. Since the responses were common to the THWELL and RPWELL we conclude that they are not (or very weakly) related to the presence of the resistive plate. More details are given below.

*Response to operation-voltage changes:* Abrupt voltage changes result in an immediate change of $E_0$ (figure 4 region b and figure 6a); it is followed by a slower change in $E_Q$ until the gain reaches stabilization. In the transition from low-to-high operation voltage (figure 6a), $E_Q$ stabilized with a characteristic time of ~1 hour. On the other hand, in the transition from high-to-low operation voltage (figure 6b) $E_Q$ stabilized with a characteristic time of several hours. This difference suggests that more than a single process is involved in the stabilization of $E_Q$; one is dominant when the gain is increased and the other when the gain is decreased.

In view of the charging up/down model we assume that the rate of the charging up process, $\varGamma_{up}$, depends on the amount of charge within the holes; the larger the charge - the larger the charging-up rate. On the other hand, the rate of the charging down, $\varGamma_{down}$, is less affected by the charge within the holes. Under these assumptions, the response for the voltage increase can be explained as follows:

- An abrupt increase of the voltage (and gain) results in an increase of $\varGamma_{up}$, while $\varGamma_{down}$ remains unchanged. The condition for a charging-up dominated profile, $\varGamma_{up} > \varGamma_{down}$, is reached.
- Charges accumulate on the insulators and $E_Q$ increases. In the cylindrical hole of the THWELL and RPWELL the outcome is a reduction of the total filed, $E$, and of the gain.
- $\varGamma_{up}$ is decreased until the condition for a steady state, $\varGamma_{up} = \varGamma_{down}$, is reached.

The response for the voltage increase can be explained in a similar manner.

*Response to irradiation-rate changes* (figure 7)*:* Abrupt changes of the irradiation rate affected $E_0$ significantly only in the presence of a resistive plate. Therefore, we distinguish between responses which are common to the THWELL and the two RPWELL electrodes, and those which are not.

- Common to all tested detectors, the stable-gain value under low irradiation rate increased after they were exposed to high irradiation rate for the first time.
- Different responses were observed to the following rate transitions:
  - The gain of the RPWELL detectors changed with every change of the irradiation rate. Typically (region "d" in figure 4), an immediate prompt change is followed by a slower component before the gain stabilizes.
  - The gain of the THWELL detector was not affected by changes of the irradiation rate.

In the following, we attempt to interpret these observations in the context of the charging-up/down model.

The different responses of the THWELL and the RPWELL to the irradiation-rate transitions, (other than the first one) demonstrate the role that the RP plays in the gain stabilization; at high rate, the current flowing through the RP results in a voltage drop (figure 4 region f). In view of the charging-up/down model, this reduces $E_0$. The drop, which could be of a few tens of volts, could explain the immediate change of the gain of the RPWELL detector with every change of the irradiation rate. Similar response was reported in previous studies with RPWELL-detectors [11-14].

The prompt change of gain is followed by a slower component. This can be attributed to changes in $E_Q$. An increase of the irradiation rate modifies accordingly the amount of charge in the holes and increases $\Gamma_{up}$. Thus, a charging-up dominated trend is expected. In the cylindrical-hole configurations, this should correspond to a gain drop until a steady state is reached. However, the gain was slowly increased (regions "d" and "f" in figure 4). This result indicates that the model of the charging- up/down of the hole walls has to be extended.

Using the model, we examined the common response of the investigated detectors to the first transition from low-to-high rate (region "d" in figure 4 and figure 7). The transition resulted in an immediate gain drop (within less than a minute); it was followed by a very slow gain increase (for over a day); the gain was stabilized at a value higher than the one measured at the low rate. When the rate was lowered again, the original gain was not restored. The gain at the steady-state was indeed stabilized at a higher value.

Since this response was common to the RPWELL and THWELL detectors, its origin cannot be related the RP, but only to changes of $E_Q$. These changes depend on $\Gamma_{up}$, which increases with the irradiation rate. Thus, it is expected to obtain a charging-up dominated trend of the gain-transient. We assume that in cylindrical-hole configurations this trend should correspond to a gain drop [9]. Therefore, the immediate gain drop in the first irradiation rate transition agrees with this description. However, the slow and long rise of the gain that followed it and the change in steady-state do not fit this picture and require further studies; they suggest either dependence on the initial conditions, i.e. charge distribution on the insulator, or an additional rate-dependent process which was not included in our assumed model.

A possible addition to the discussed model could be that charges accumulate also on the rim of the THWELL or RPWELL top electrode. It could be that an increase of the irradiation rate increases significantly the electron charging-up of the THGEM's top rim. The additional charge distribution on the rim would result in an electric field with opposite orientation to that induced by the charge accumulating on the hole walls. Thus, reducing its effect and increasing the gain. Moreover, it could be that $\Gamma_{down}$ in the rim area is very low. This could be the cause for the change of the stable-gain value at lower irradiation rates. Such effect is not expected in the case of an increase in the operation voltage, which, in THWELL and RPWELL structures, modifies the charge only inside the holes. This assumption is supported by the observation reported in [10]. There, it is mentioned that a slow and long rise in the gain was observed only in electrodes with rims. Even though that the used $^{55}$Fe x-ray source had low rate, the main contribution to the increase of the gain could have been from the charging-up of the THGEM's rims by either the avalanche-electrons or ions accumulation.

Future work should focus on understanding more deeply the response of these detectors to rate transitions. Monte Carlo simulations incorporating the charging up/down model could be useful.

**Acknowledgments**

We are thankful to L. Arazi, M. Pitt, L. Moleri, A.E.C. Coimbra, P. Bhattacharya and E. Erdal, of our Research Group, for useful discussions. This research was supported in part by the I-CORE Program of the Planning and Budgeting Committee, the Nella and Leon Benoziyo Center for High Energy Physics, the Mel and Joyce Eisenberg-Keefer Fund for New Scientists and by a Grant No 712482 from the Israeli Science Foundation (ISF).